\documentclass[fleqn,usenatbib]{mnras}

\usepackage[T1]{fontenc}
\usepackage{ae,aecompl}

\usepackage{graphicx}	
\graphicspath{{figures/}} 

\usepackage{amsmath}	
\usepackage{amssymb}	
\usepackage{bm} 
\usepackage{balance}
\usepackage{siunitx}

\usepackage[usenames,dvipsnames]{xcolor}
\usepackage{verbatim}

\def\be{\begin{equation}}
\def\ee{\end{equation}}
\def\bea{\begin{eqnarray}}
\def\eea{\end{eqnarray}}

\providecommand{\sorthelp}[1]{}
\hypersetup{
    colorlinks = true,
    citecolor = {MidnightBlue},
    linkcolor = {BrickRed},
    urlcolor = {BrickRed}
}

\usepackage{geometry}
\title[Characterizing line-of-sight variability of polarized dust emission]{Characterizing line-of-sight variability of polarized dust emission with future CMB experiments}

\author[Lisa McBride et al.]{
Lisa McBride,$^{1,2}$\thanks{E-mail: lisaleemcb@gmail.com}
Philip Bull,$^{3,4,5}$
Brandon S. Hensley$^{6}$
\\
\\
$^{1}$Department of Physics and McGill Space Institute, McGill University, Montreal, QC H3A 2T8, Canada\\
$^{2}$Department of Physics \& Astronomy, San Francisco State University, San Francisco, CA 94132, USA\\
$^{3}$Jodrell Bank Centre for Astrophysics, University of Manchester, Manchester, M13 9PL, United Kingdom \\
$^{4}$Department of Physics and Astronomy, University of Western Cape, Cape Town 7535, South Africa \\
$^{5}$Astronomy Unit, Queen Mary University of London, London E1 4NS, United Kingdom \\
$^{6}$Department of Astrophysical Sciences,  Princeton University, Princeton, NJ 08544, USA
}

\date{Accepted XXX. Received YYY; in original form ZZZ}

\pubyear{2022}

\begin{document}
\label{firstpage}
\pagerange{\pageref{firstpage}--\pageref{lastpage}}
\maketitle

\begin{abstract}
 Galactic dust emission is often accounted for in cosmic microwave background (CMB) analyses by fitting a simple two-parameter modified blackbody (MBB) model in each pixel, which nominally accounts for the temperature and opacity of the dust. While this may be a good approximation for individual dust clouds, typically a number of such clouds are found along each line of sight and within each angular pixel, resulting in a superposition of their spectra. In this paper, we study the effects of this superposition on pixel-based foreground fitting strategies by modelling the spectral energy distribution (SED) in each pixel as the integral of individual MBB spectra over various physically-motivated statistical distributions of dust cloud properties. We show that fitting these SEDs with the simple two-parameter MBB model generally results in unbiased estimates of the CMB Stokes Q and U amplitudes in each pixel, unless there are significant changes in both the dust SED and polarization angle along the line of sight, in which case significant ($ > 10\sigma$) biases are observed in an illustrative model. 
We also find that the best-fit values of the dust temperature, $T_d$, and spectral index, $\beta_d$, are significantly biased away from the mean/median of the corresponding statistical distributions when the distributions are broad, suggesting that MBB model fits can give an unrepresentative picture of the physical properties of the dust at microwave wavelengths if not interpreted carefully. Using a Fisher matrix analysis, we also determine the experimental sensitivity required to recover the parameters of the $T_d$ and $\beta_d$ distributions themselves by fitting a probabilistic MBB model, finding that only the parameters of broad distributions can be measured by SED fitting on a single line of sight.
\end{abstract}


\begin{keywords}
cosmic background radiation -- polarization -- dust, extinction
\end{keywords}



\section{Introduction} \label{sec:intro}
As cosmic microwave background (CMB) polarization experiments rapidly gain in sensitivity, it is becoming increasingly important to build a detailed understanding of Galactic dust emission. As well as being the brightest component of the polarized microwave sky at frequencies above $\sim$100\,GHz, dust emission is also one of the most complex, with evidence of significant spectral and spatial structure in polarization \citep{Planck_frequency, PlanckDust, Guillet_2018, pelgrims2021evidence, Ritacco2022} that is non-trivial to model and remove \citep[e.g.,][]{BICEP_Keck, kogut2016foreground, Planck_foregrounds, Hensley_2018, mangilli2019dust}. This is concerning for the low amplitude, large angular scale phenomena that many CMB polarization experiments are targeting, particularly the B-mode signal from primordial gravitational waves that has the potential to constrain fundamental physics and provide concrete physical constraints on an inflationary period in the early Universe \citep{2016ARA&A..54..227K}. 
Upcoming experiments like Simons Observatory \citep{SimonsObs}, LiteBIRD \citep{LITEBIRD}, and CMB-S4 \citep{cmb-s4} aim to make a measurement of the tensor to scalar ratio, $r$, with an uncertainty on the order of $\sigma(r) \sim 10^{-3}$ or below. In order to achieve this, residual foreground power must be reduced to below the level of $D^{\rm BB}_\ell \sim 10^{-5} {\mu{\rm K}}^2$ in the BB power spectrum on scales $\ell \lesssim 100$ \citep{cmb-s4, Remazeilles_2016}.

A common approach has been to assume a simple empirical model for the thermal dust emission, the modified blackbody (MBB), which takes a thermal (blackbody) spectrum characterized by the dust temperature, $T_d$, and modifies it with an opacity factor that scales with frequency as $\nu^{\,\beta_d}$. While employing only two spectral parameters, plus an amplitude per pixel, this model has proven adequate for modeling both total and polarized dust emission at the sensitivities of current CMB experiments both at the map level \citep{Planck_foregrounds} and power spectrum level \citep{BICEP2018,Planck_2018_XI}.

A question that immediately arises is whether the simple MBB fitting approach gives rise to model errors that can bias the recovered CMB signal when applied to the real sky at the sensitivities of next-generation experiments. The physical properties of interstellar dust grains, such as temperature and composition, are known to vary in the interstellar medium (ISM), resulting in spatial variability of the dust spectral energy distribution (SED). Indeed, analysis of component-separated maps of the microwave sky has revealed variations of the dust SED on the $\gtrsim 1^\circ$ scales relevant to CMB B-mode analyses \citep{Planck_foregrounds, Planck_2018_XI, Ritacco2022}. These variations give rise to ``frequency decorrelation,'' i.e. the map of dust emission at one frequency is not simply equivalent to the map at another frequency rescaled by a spatially-constant frequency-dependent multiplicative factor. The level of frequency decorrelation between microwave frequencies is a major uncertainty in current CMB analyses \citep{BICEP2018, S4_Forecasting}.

Just as dust properties can vary across the sky, they can also vary along the line of sight. If a sightline intersects clouds with different dust SEDs, e.g., due to the dust having different compositions or temperatures, and if the magnetic fields in these clouds are misaligned, then the polarization angle of the dust emission becomes decorrelated across frequencies \citep{tassis2015searching}. This ``line of sight frequency decorrelation'' is a clear indicator of spatially variable dust SEDs, and has recently been detected in Planck data \citep{pelgrims2021evidence}. In order to produce the observed level of line of sight frequency decorrelation, \citet{pelgrims2021evidence} found that the shape of the polarized dust SED between 217 and 353\,GHz must vary at the $\simeq 10\%$ level from cloud to cloud.

In addition to complexity induced by spatial variability, dust emission even from a single grain population localized along the line of sight is unlikely to correspond perfectly to a MBB. Detailed physical models of dust polarization have been constructed based on derived material properties of interstellar grains \citep[e.g.,][]{Draine_2009, Guillet_2018, Draine2021ApJ}. The models elucidate how the polarized dust SED changes in response to different grain size distributions, grain shapes, grain porosities, dust alignment properties, the intensity and spectrum of the ambient radiation field, and the relative abundances of different grain materials. These models are described by a large number of parameters connected to the physics of dust and the interstellar environments in which dust resides, each of which may vary both across the sky and along the line of sight. 

Given these complications, but also lack of evidence for inadequacy of the MBB in current data, it is imperative to understand whether and how the MBB parameterization could lead to biases when used in CMB data analyses. It has already been demonstrated that measurements can indeed be biased if more complicated underlying dust models are assumed \citep[e.g.,][]{kogut2016foreground, Remazeilles_2016, Hensley_2018, Errard2022}. Even assuming a modest `2MBB' extension, where a single MBB model is replaced by the combination of two separate MBB signals, can cause a significant bias in the recovered tensor to scalar ratio measurement \citep{Remazeilles_2016}.

In this paper, we use parametric MBB model fitting as a convenient reference case. Our aims are to study not only possible biases in recovery of polarized CMB information, but also the recovery of the dust model parameters themselves. In particular, we quantify whether MBB fits are capable of reliably recovering summary information about more complex true distributions of dust cloud properties along each line of sight, or whether physical conclusions drawn from such model fits may be misleading. We do this by studying a range of more complex dust SEDs for individual lines of sight, constructed by integrating simple MBB models over various probability distributions with physically motivated properties, including one that allows line-of-sight variations in polarization angle that can lead to more complex spectral structure. For simplicity, we consider {\it only} the Stokes Q and U parameters, which allows us to avoid making additional model assumptions about the unpolarized fraction of the dust emission.

The paper is organized as follows. In Section~\ref{sec:models}, we present the set of dust models used in our analysis, and we set out our single-pixel simulation and parameter-fitting methodology in Section~\ref{sec:recovery}.
We investigate the implications of each dust model scenario in Section~\ref{sec:results}, both for producing biased model fits, and to learn how well the statistical parameters of the model can be constrained from observations, and then summarize our results in Section~\ref{sec:conclusions}.

\section{Line-of-sight dust SED models} \label{sec:models}
In this section, we outline the various dust models used in our analysis. These include a simple single modified blackbody (sMBB) model; a generic `probabilistic' (pMBB) model based on integrating the basic MBB SED over a distribution of model parameters; and a Turbulent Magnetic Field model (TMFM) based on integrating a varying polarization angle over the line-of-sight. These models are then used, in conjunction with a synchrotron and CMB model, to generate our simulated data.

\begin{figure}
\includegraphics[width=\columnwidth]{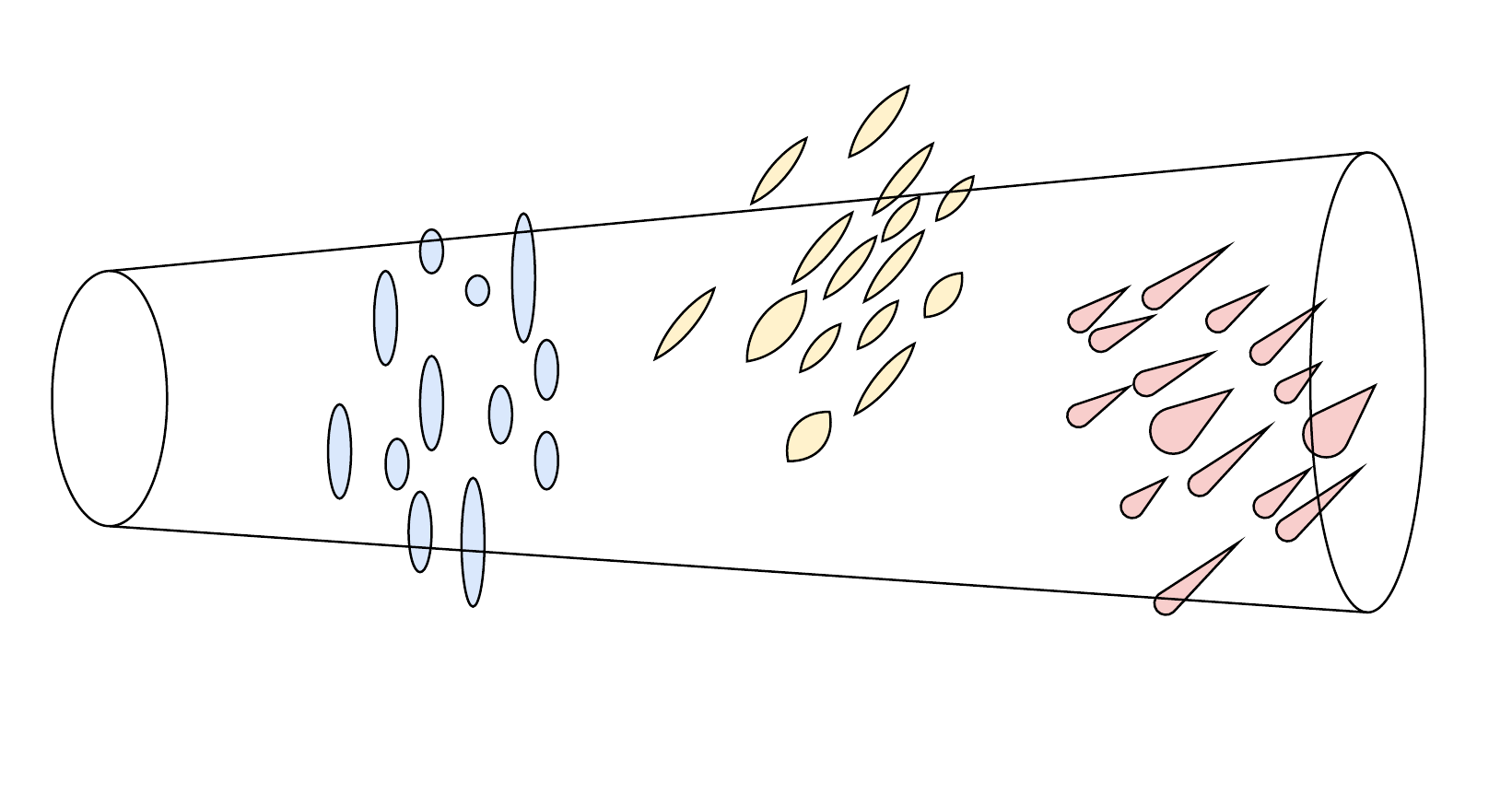}
\caption{A simplified picture of dust populations along a single line-of-sight with angular extent. Polarized emission from dust is due to alignment of non-spherical grains with the Galactic magnetic field. Different dust clouds sample different regions along the line of sight and therefore could have different magnetic field orientations. The composition and temperature of dust in each cloud can also differ. All of the emission within the conic column is collapsed into a single combined SED, which is measured at the antenna.}
\label{fig:dust_picture}
\end{figure}

\subsection{Single-population dust models (sMBB)} \label{subsec:single-pop}

In the optically-thin limit, thermal dust emission at CMB frequencies has been shown empirically \citep{Planck_foregrounds, Planck2018XI} to be well-described by a modified blackbody of the form 
\be \label{eq:smbb}
I_{\nu}^{\text{MBB}} = A_d^i \bigg( \frac{\nu}{\nu_0^d} \bigg)^{\beta_d} B_{\nu}(T_d),
\ee
where the function $B_{\nu}(T)$ is the Planck function with temperature $T_d$, and the power-law term represents a frequency-dependent opacity with spectral index $\beta_d$. In terms of the physical interpretation of this model, $T_d$ represents an assumed uniform effective temperature for the dust, and $\beta_d$ is determined by its composition. A final parameter, $A_d^i$, is the intensity of the emission for polarization $i$ at the reference frequency, $\nu^d_0 = 353$\,GHz.

As discussed in the previous section, we adopt this `sMBB' model as our reference model. For our analysis, the amplitude parameters are chosen such that the amplitude at the reference frequency is consistent the observed high Galactic latitude values from the \texttt{Commander} analysis in \cite{Planck_foregrounds}, i.e., $A^Q_d = A^U_d = 3.5 \,\mu\text{K}_{\text{RJ}}$. We adopt fiducial values of ${T}_d = 20 \text{ K}$ and ${\beta}_d = 1.6$, which are also broadly representative of the Planck data \citep{Planck_2018_XI}. We ignore any polarization effects that could lead to a different frequency dependence of the Q versus U polarization, only relaxing this assumption for the TMFM model in Sect.~\ref{subsec: TMFM}.

\subsection{Probabilistic models} \label{subsec:stat_models}

The observed dust emission in an angular pixel on the sky is an integral of the flux along the line-of-sight, which based on observations including the distribution of neutral hydrogen, is believed to encompass different physical dust emission regions \citep[i.e., `clouds',][]{Panopoulou2020ApJ, Clark2019ApJ, pelgrims2021evidence}. It is physically unlikely that the dust emission will be uniform along any given line of sight; instead, there will be variability both between, and within, different regions. The physical properties of each cloud, as well as the distribution from which their properties are drawn, are currently unknowns. Constraints on these quantities would be of tremendous interest to efforts to characterize the ISM.

For the purposes of our model, we assume that thermal dust emission can be described, in generality, by a single model, $I^{\text{model}}_{\nu}(\vec{\theta})$, with only the values of the model parameters $\vec{\theta}$ varying both internally, and from region to region (see Fig.~\ref{fig:dust_picture}). The possible values of these parameters are then described by a continuous statistical distribution under the assumption that along a single LOS there is sufficient variability both between regions, and within individual clouds themselves, such that the entire statistical distribution is well-sampled. The resulting model for the intensity along a single line of sight is
\be \label{eq:probmbb}
I_{\nu} (\vec{\sigma}) = \int I^{\text{model}}_{\nu} (\vec{\theta}) \, p(\vec{\theta}; \vec{\sigma})\, d\vec{\theta},
\ee
where $p(\vec{\theta}; \vec{\sigma})$ is the probability distribution function (pdf) for the parameters, which itself is parametrized by a set of hyperparameters $\vec{\sigma}$ that describe its shape. The pdf also carries an implicit normalization which is subsumed into the overall amplitude parameter.

\begin{table}
\caption{Fiducial parameter values for the foreground models used in our analysis. For the statistical dust models, hyperparameters of the Gaussian spectral parameter distributions are specified, while the sMBB and synchrotron models are deterministic. The reference frequencies for all models are $\nu_{\rm ref} = 353$~GHz, except synchrotron, which has $\nu_{\rm ref} = 30$~GHz.}
\centering 

\hspace{-1em}\begin{tabular}{c c c c c c c c } 
\hline & \\[-2ex] 
Component  & $\overline{\beta}$ & $\overline{T}$ [K] & $\sigma_{\beta}$ & $\sigma_T$ [K] & $\kappa$ \\ [0.8ex] 
\hline 
sMBB  & \,1.6 & 20 & -- & -- & -- \\ 
\hline
pMBB narrow  & \,1.6 & 20 & 0.02 & 0.4 & -- \\
pMBB inter.  & \,1.6 & 20 & 0.10 & 2.0 & -- \\
pMBB broad  & \,1.6 & 20 & 0.20  & 4.0 & -- \\
\hline
TMFM narrow  & \,1.6 & 20 & 0.02 & - & 0.1 \\
TMFM inter.  & \,1.6 & 20 & 0.10 & - & 1 \\
TMFM broad  & \,1.6 & 20 & 0.20 & - & 100 \\
\hline
Synchrotron  & $-$1.2~ & -- & -- & -- & -- \\ 
\hline 
\end{tabular}
\label{table:distributions} 
\end{table}

\subsubsection{Probabilistic MBB model (pMBB)} \label{subsec:pMBB}

We now consider a specific case of the statistical model described above. We assume a functional form for the model SED that is given by a single MBB model (Eq.~\ref{eq:smbb}), and a joint pdf for the dust temperature and spectral index parameters that is an uncorrelated Gaussian distribution,
\be
p(\beta_d, T_d; \overline{T}_d, \bar{\beta}_d, \sigma_{T}, \sigma_{\beta})
\propto e^{ {-\frac{(T_d - \overline{T}_d)^2}{2\sigma_T^2}} }
 e^{{-\frac{(\beta_d - \bar{\beta}_d)^2}{2 \sigma_{\beta}^2}} } \nonumber
\ee
where $\overline{T}_d$ and $\bar{\beta}_d$ are the means of the dust temperature and spectral index pdfs, and $\sigma_{T}$ and $\sigma_{\beta}$ are the standard deviations. The Gaussian normalisations and all other normalising factors are subsumed into an overall amplitude factor per polarization, $A^i_d$. We call this the probabilistic MBB (pMBB) model.

For the fiducial values of the hyperparameters, we again use the Planck results as our guide. Mean values of $\overline{T}_d = 20 \text{ K}$ and $\bar{\beta}_d = 1.6$ were chosen to be consistent with the sMBB model parameters (see Sect.~\ref{subsec:single-pop} and Table~\ref{table:distributions}), such that, in the limit that the variance of the pdf goes to zero, the sMBB model is recovered.

\begin{figure*}
\includegraphics[width=1.8\columnwidth]{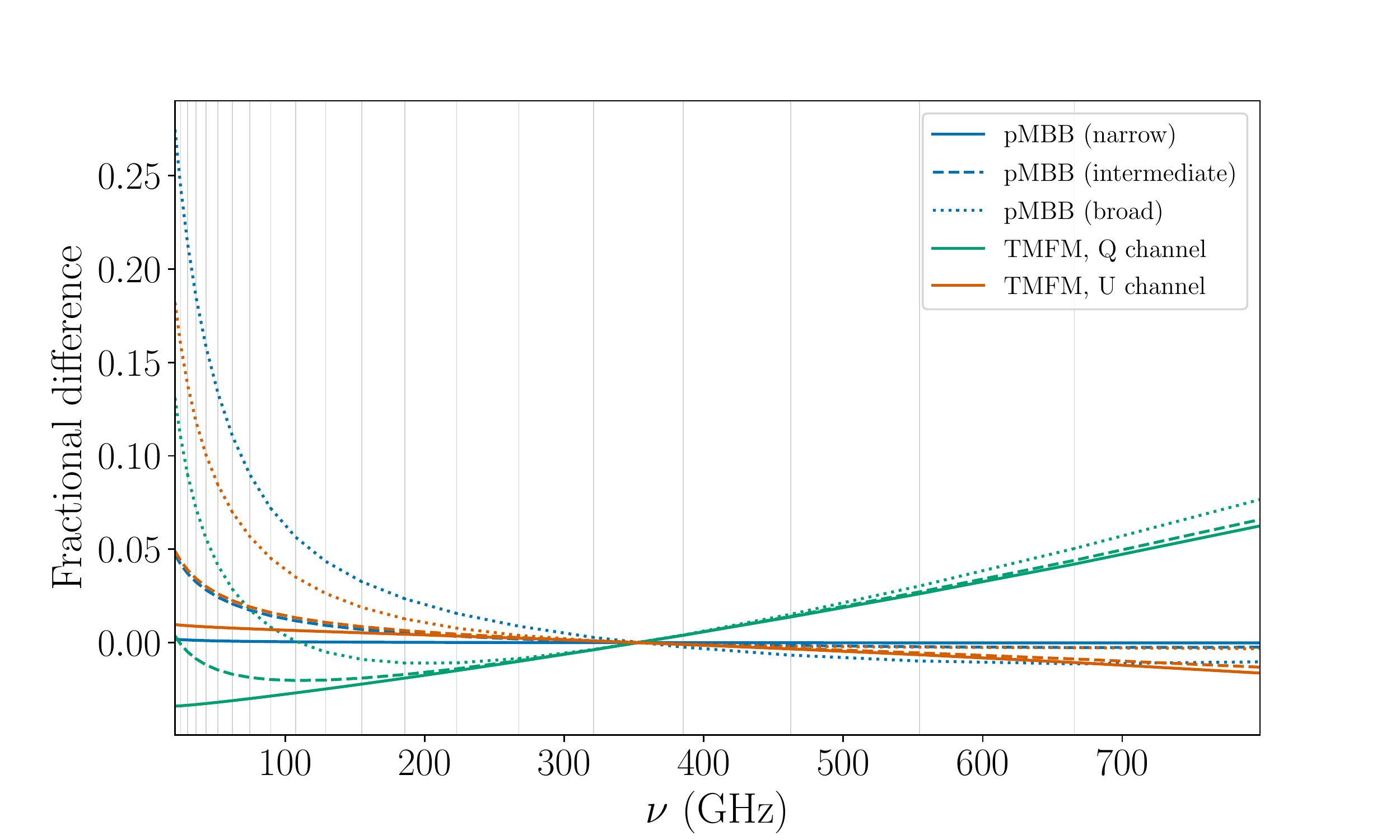}
\caption{Fractional difference in dust SEDs between each input model and a base model (sMBB), divided by the base model, for probabilistic MBB models with broad, intermediate, and narrow temperature and spectral index distributions are shown in blue (see Table~\ref{table:distributions} for the explicit values of the statistical parameters). Equivalent cases for the TMFM model are shown separately for the Stokes Q (green) and U (orange) channels. Vertical grey lines correspond to the frequency bands of the PICO mission \citep{hanany2019pico}.}
\label{fig:frac_diff}
\end{figure*}

For the standard deviation parameters, three illustrative cases were considered, corresponding to narrow, intermediate, and broad spreads in the distribution of possible values of the physical parameters. The values taken for each case are summarized in Table~\ref{table:distributions}. These values are roughly consistent with the dispersion between pixels, as derived by \citet{Planck2015XLVIII}. We plot the fractional difference of the SED for each case with that of the sMBB model in Fig.~\ref{fig:frac_diff}.

For completeness, we also considered whether alternative forms for the pdf (e.g., with heavier tails, or skewness) could substantially affect the SED. Replacement of the Gaussian pdf with Lognormal and Gamma pdfs did yield slightly different SEDs, but on fitting single MBB models to each model and fixing the pdfs to the same mean and variance, the resulting best-fit values of the sMBB parameters were very similar. As such, we do not consider the effects of differing pdfs further here, restricting our attention solely to the Gaussian case.

\subsubsection{Turbulent Magnetic Field model (TMFM)} \label{subsec: TMFM}

In the above set of pMBB models we have ignored any polarization-dependent effects on the dust SEDs. Different regions along the line of sight can have different dust physical properties {\it and} magnetic field orientations. While the emission from each region simply adds in total intensity, they add vectorially in polarization. This leads to preferential amplifications and cancellations of the signal at some frequencies, imparting additional spectral structure in the Stokes Q and U channels compared with total intensity \citep{tassis2015searching, PohDodelson, pelgrims2021evidence}. 
As a result, we can no longer treat the shape of the SED as being the same in Stokes Q as in Stokes U, necessitating the addition of more spectral model parameters. The shapes of the polarized SEDs also become position-dependent, with differences in magnetic fields along different lines of sight causing spatial variations in the resulting SEDs, and cancellations along the line of sight also reduce the overall polarized fraction.

\begin{figure}
\centering
\includegraphics[width=1.02\columnwidth]{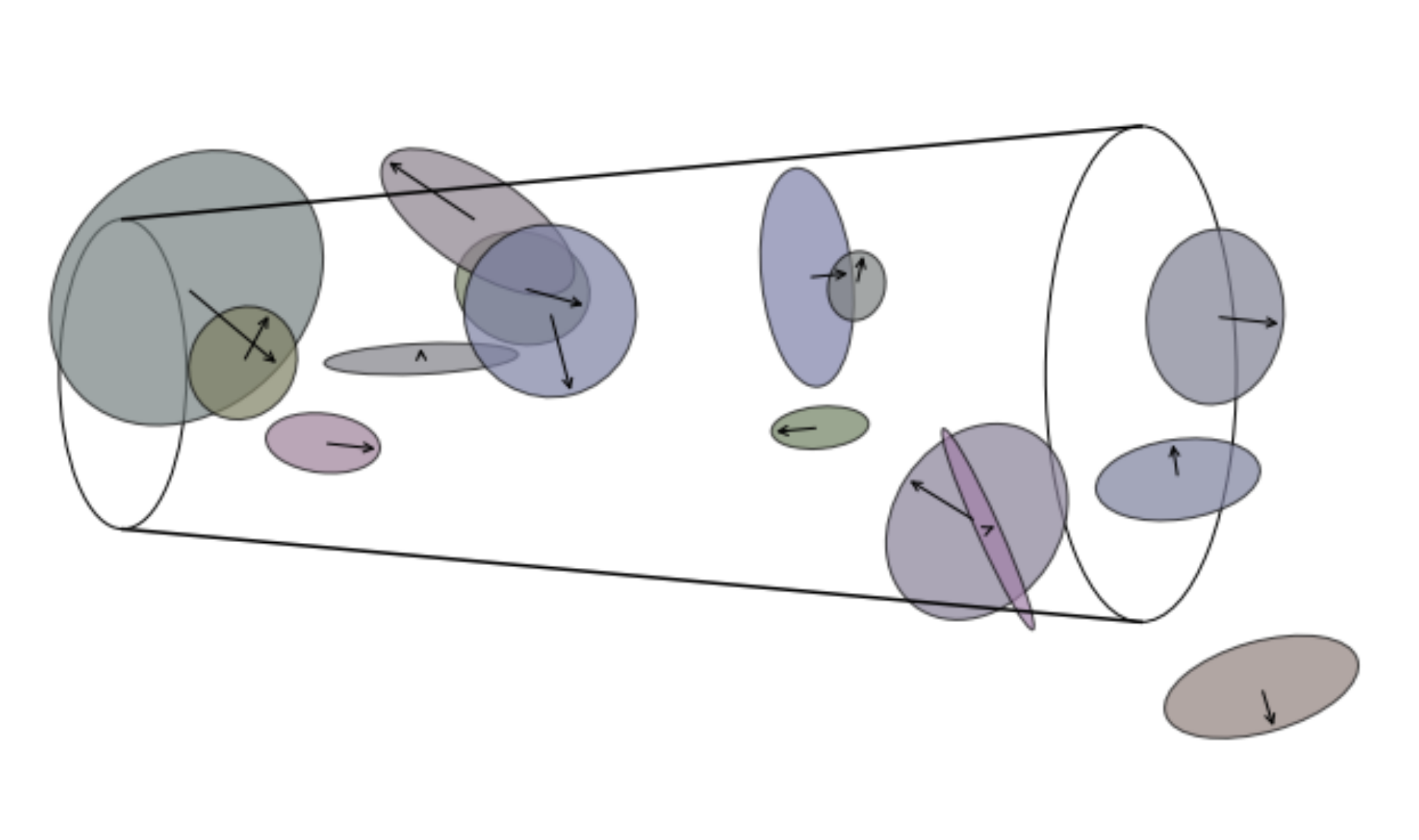}
\vspace{-1em}
\caption{A simplified picture of the Turbulent Magnetic Field model (TMFM) for a single line of sight. In this model, dust clouds are allowed to have different polarization angles $\chi$, which are randomly drawn from a von Mises distribution, and are assumed uncorrelated. The vectorial nature of the polarization signal leads to variations in the SEDs between the Stokes Q and U channels.}
\label{fig:TMFM_picture}
\end{figure}

To investigate the potential impact of these `decorrelation' effects on the fitting process, we construct a simple model in which there is a distribution of polarization angles along each line-of-sight. We call this the Turbulent Magnetic Field model (TMFM), as it describes a scenario in which the Galactic magnetic field structure is dominated by approximately uncorrelated random fluctuations in orientation on small scales (Fig.~\ref{fig:TMFM_picture}).
This is consistent with the phenomenological model used for analysis in \citet{Planck2018XII}, and agrees with the relationships between various polarization statistics like the polarization fraction and polarization angle dispersion.

For the distribution of polarization angles $\chi$, we assume a von Mises distribution, $P(\chi;\, \bar{\chi}, \kappa)$, with mean angle $\bar{\chi}$ and shape parameter $\kappa$ that governs the width of the distribution. The von Mises distribution is an analogue of the Gaussian distribution on the unit circle. Since dust at different orientations is expected to come from physically distinct regions, we model the dust temperature in each region, $T_d$, as a simple function of the polarisation angle, $\chi$,
\begin{equation} \label{eq:Tdust}
T_d\left(\chi\right) = T^\prime_d + \frac{T_\chi\, \Delta\chi}{\pi/2}~~~,
\end{equation}
where $T^\prime_d$ is a fixed reference dust temperature, and $T_{\chi}=4\,{\rm K}$ defines the maximum contribution to the effective dust temperature due to the alignment of the polarised emission. Lastly, $\Delta\chi \in [0, \pi/2)$ is the difference between $\chi$ and $\bar{\chi}$. This is a convenient toy model used to illustrate the physical effects of correlation between the polarized dust SED and the polarization angle; we do not intend to suggest that there is any causal relationship between the magnetic field orientation and the dust temperature.

In the following analysis, the reference dust temperature is set to the previous mean of the pMBB models, $T^\prime_d=20 \, K$. The intensities for the Q and U channels are then defined via integrals over the von Mises distribution for $\chi$,
\begin{eqnarray}
Q_\nu^\prime(\beta) &= \left(\frac{\nu}{\nu_0}\right)^\beta \int B_\nu\left(T_d\left(\chi\right)\right) \cos(2\chi) \frac{dP}{d\chi} d\chi \\
U_\nu^\prime(\beta) &= \left(\frac{\nu}{\nu_0}\right)^\beta \int B_\nu\left(T_d\left(\chi\right)\right) \sin(2\chi) \frac{dP}{d\chi} d\chi.
\end{eqnarray}
As in all of the other cases, we assume that the Q and U amplitudes are equal, resulting in a mean angle, $\bar{\chi} = \pi / 8$.
In contrast to the pMBB model, the parameter model with a statistical distribution is the polarization angle of the dust emission in each cloud rather than its dust temperature (although the dust temperature also fluctuates due to its dependence on $\chi$). We again assume that the dust temperature is uncorrelated with the opacity parameter, $\beta_d$, which as in our previous models follows a Gaussian distribution. The final SED for this model is then
\be
Q_\nu = A_Q \int Q_\nu^\prime(\beta) \exp \left ({-\frac{(\beta_d - \bar{\beta}_d)^2}{2 \sigma_{\beta}^2}} \right )\, d\beta,
\ee
with a similar expression holding for $U_\nu$. In what follows, we will consider three different values of the von Mises shape parameter, $\kappa$, to illustrate the effects of line of sight frequency decorrelation (see Table~\ref{table:distributions}).

\subsection{Other signal components}

The microwave sky is a superposition of many radiation sources, including synchrotron emission, anomalous microwave emission (AME), free-free emission, and the CMB in addition to Galactic dust. Since our focus is only on the polarized emission in this paper, we neglect two of these sources: free-free (bremsstrahlung) emission, which is mostly present at low Galactic latitudes and is only weakly polarized,
and AME, for which there is a stringent empirical upper bound of $\sim 2\%$ on its polarization fraction \citep{kogut2007three, Planck_frequency, macellari2011galactic, Herman2022} and theoretical arguments predicting negligible polarization \citep{draine2016quantum}.

With this in mind, we consider a sky signal composed only of the CMB signal, synchrotron emission, and thermal dust emission in our analysis. We adopt values of the synchrotron signal to be consistent with the Planck \texttt{Commander} results at high Galactic latitudes, i.e., polarization amplitudes of $A^Q_s = A^U_s = 280\, {\rm Jy}\, {\rm sr}^{-1}$ at a reference frequency of 30\,GHz. We also set the synchrotron spectral index to $\beta_s = -1.2$ in units of flux density. 
For the CMB polarization signal, we adopt representative values of $Q_{\nu} = U_{\nu} = 0.6 \, \mu \text{K}_{\text{CMB}}$. In all models, we assume a constant polarization angle for both the CMB and synchrotron emission, the latter agreeing with the dust polarization angle (or, in the case of the TMFM, the mean dust polarization angle). An overview of the parameters used for each foreground model is given in Table~\ref{table:distributions}.

\section{Recovering dust properties from SED fits} \label{sec:recovery}

In this Section we describe the noise properties of our single-pixel simulations and give an overview of the single-pixel model fitting procedure that was previously presented in \cite{Hensley_2018}.

\subsection{Single-pixel simulations} \label{subsec:sim_data}
Simulated data vectors of the Stokes Q and U intensity per pixel were generated using the \texttt{SinglePixel}\footnote{\url{http://philbull.com/singlepixel/}} package. Each data vector represents the intensity along a single line-of-sight for a set of frequency bands, assuming a common angular smoothing of $1^{\circ}$ FWHM for all bands, which is appropriate for a CMB B-mode analysis.

We adopt the noise properties and band specification of PICO \citep{sutin2018pico, hanany2019pico}, a proposed space-based polarimeter intended to conduct full-sky surveys with a few arcmin. resolution in each band across 21 bands between 21 and 799\,GHz. We assumed a delta function bandpass at each center frequency, and added uncorrelated Gaussian noise according to the noise rms per frequency band specified in \cite{sutin2018pico} (and adjusting to 1 deg$^2$ pixels). Each data vector contains contributions from the CMB, synchrotron, and a particular thermal dust emission model. As discussed above, we neglected the Stokes I channel.

In what follows, data vectors constructed with the single MBB model for the thermal dust will be referred to as `sMBB data', and similarly for the other dust models. We include the same CMB and synchrotron components in all data vectors, and generate $200$ noise realisations for each dust model. The explicit values for all of the input models are presented in Table~\ref{table:distributions}.

\subsection{MCMC-based model fitting procedure} \label{subsec:fitting}

We constructed a simple Gaussian likelihood for the simulated data, using the input noise covariance per frequency band (which assumes no correlations between bands), and assuming only a single pixel per dataset. Uniform priors are assumed for all parameters; see Table~\ref{table:priors} for the corresponding ranges. Note that these priors will be informative for some parameters; alternatively a Jeffreys prior could be adopted for (e.g.) the spectral index parameters \citep{Eriksen:2007mx, 2019MNRAS.490.2958J}, but we do not study this question further here.
We then used the \texttt{emcee} affine-invariant ensemble sampler \citep{Foreman_Mackey_2013} to sample from the joint posterior distribution of the CMB amplitude parameters, and amplitude and spectral parameters for the synchrotron component and a chosen dust model. We initialized 48 walkers per run, allowing 1000 steps of burn-in before running each walker for 10,000 steps. Note that in all cases considered in this paper, we fit only to the Stokes Q and U data, and always include synchrotron and CMB models with free Q and U amplitude parameters, and free spectral parameter $\beta_s$ for the synchrotron component.

\begin{table}
\caption{Uniform prior ranges used for the MCMC fitting procedure.} 
\centering 

\def\arraystretch{1.2}
\begin{tabular}{c c l} 
\hline\hline 
Parameter & Notation &  Prior range \\ [0.8ex] 
\hline 
sMBB amplitudes Q, U  & $A^{Q,U}_{\rm{sMBB}}$ & [1,100]~$\mu$K  \\ 
pMBB amplitudes Q, U & $A^{Q,U}_{\rm{pMBB}}$ & [1,100]~$\mu$K  \\
Dust temperature & $T_d$ & [16, 24]~K \\
Dust spectral index & $\beta_d$ & [1.4, 1.8] \\
Dust temperature std. dev. & $\sigma_T$  & [0.1, 10]~K \\ 
Dust spectral index std. dev. & $\sigma_{\beta}$  & [0.01, 1] \\ 
Synchrotron spectral index & $\beta_s$  & [-1.6, -0.8] \\ [1ex] 
\hline 
\end{tabular}
\label{table:priors} 
\end{table}

\begin{table*}
\centering 
\def\arraystretch{1.4} 
\begin{tabular}{ l || S[table-format=3.2] S[table-format=3.2] S[table-format=3.2] S[table-format=3.2] S[table-format=3.2] S[table-format=3.2] S[table-format=3.2] S[table-format=3.2] S[table-format=3.2]}
 \hline
 \hline
{\bf Dust model} & {$A_{\text{MBB}}^{Q}$} & {$A_{\text{MBB}}^{U}$} & {$A_{\text{CMB}}^{Q}$} & {$A_{\text{CMB}}^{U}$} & {$A_{\text{S}}^{Q}$} & {$A_{\text{S}}^{U}$} & {$\beta_d$} & {$T_d$} & {$\beta_S$} \\
 \hline
 sMBB  & -0.11 & -0.10 & -0.02 & -0.01 & 0.01 &  0.11 & 0.17 & -0.17 & 0.03 \\
 \hline
 pMBB narrow  & 0.03 & -0.0041 & 0.11 & -0.07 & -0.04 & -0.023 & -0.15 & 0.14 & -0.07 \\
 pMBB intermediate  & 0.11 & 0.13 & -0.10 & -0.17 & 0.19 & 0.13 & -2.7 & 2.2 & 0.25 \\
 pMBB broad  & 0.19 & 0.22 & 0.037 & 0.070 & 0.19 & 0.27 & -12 & 9.9 & 0.037 \\
 \hline
 TMFM narrow  & 41 & -46 & -14 & 16 & 4.1 & -4.5 & 0.18 & 7.1 & -0.054  \\
 TMFM intermediate  & 41 & -46 & -14 & 17 & 3.9 & -4.6 & -0.52 & 8.9 & -0.023 \\
 TMFM broad & 40 & -47 & -0.0015 & 18 & 3.7 & -5.0 & -2.5 & 14 & -0.51 \\
 \hline
 sMBB (decoupled)  & 0.039 & -0.031 & -0.079 & -0.020 & -0.026 & 0.57 & -0.081 & -0.069 & -0.0029 \\
 pMBB broad (decoupled)  & 0.16 & 0.14 & 0.063 & 0.00090 & 0.21 & 0.23 & -0.085 & -0.071 & -0.0030 \\
 TMFM broad (decoupled)  & -0.77 & -1.1 & 1.5 & 1.8 & -0.31 & -0.33 & -0.086 & -0.054 & 0.0045 \\
 \hline
\end{tabular}
\caption{Mean error-normalized bias, $\langle \Delta \theta / \sigma_\theta \rangle$, for all free model parameters in each scenario, where an sMBB + CMB + synchrotron model is being fitted to Stokes Q and U data in all cases. Absolute values of this quantity greater than $\sim 1$ denote large biases.} 
\label{table:biases} 
\end{table*}

The fitting procedure was repeated for each of 200 noise realizations for each of the cases we considered. We then calculated summary statistics such as the marginal median and standard deviation for each parameter from the posterior for each run, and used these to calculate the error-normalized bias,
\be \label{eq:errnormbias}
\frac{\Delta \theta}{\sigma_{\theta}} = \frac{\theta_{\text{fit}} - \bar{\theta}}{\sigma_{\theta}},
\ee
where $\theta_{\text{fit}}$ is the best-fit value from the MCMC (the median of the marginal posterior in all cases considered here), and $\bar{\theta}$ is the input parameter value, or the mean of the input Gaussian distribution that was used to generate the simulated signal. In this context, $\sigma_{\theta}$ is the marginal standard deviation calculated from the posterior for each noise realization, found by determining the 68\% confidence interval for the parameter.

The error-normalized bias quantifies the level of discrepancy between the recovered best-fit value of a parameter and the input value to the simulation. It is similar to a `Z-score' or `standard score' except that the normalizing factor is the standard deviation calculated from the posterior rather than the population standard deviation. In the cases where $\bar{\theta}$ is the mean of an input parameter distribution, e.g., for the pMBB models, this quantity gives an indication of how well the best-fit model parameters represent the mean of that distribution.

If the error-normalized bias differs systematically from zero over many realizations, this is an indication that the best-fit model parameters are either biased or not representative of the properties of the input parameter distribution, potentially leading to flawed physical inferences. A simple measure of this is the mean error-normalized bias, $\langle {\Delta \theta}/{\sigma_{\theta}} \rangle$, where the angle brackets denote averaging over the ensemble of runs with different noise realizations.

\section{Results} \label{sec:results}

In this section we consider three sets of scenarios: fitting an sMBB model to the data when the underlying dust model is actually a more complex probabilistic MBB model (Sect.~\ref{subsec:pmbb}); the same, but for an underlying TMFM model (Sect.~\ref{subsec:TMFM}); and recovering the pMBB model hyperparameters from the data to infer the dust parameter distributions themselves (Sect.~\ref{subsec:hyperparams}). A summary of results for the parameter biases in each case is shown in Table~\ref{table:biases}.

\subsection{Fitting the sMBB model to pMBB data}
\label{subsec:pmbb}

In this section, we consider data vectors that have been generated using each of the pMBB models, but which we fit using an sMBB model. The key questions that we would like to answer are whether recovery of the polarized CMB can be biased as a result of the greater complexity of the underlying dust SEDs, and whether the best-fit sMBB model parameters are representative of the average properties of the population of dust clouds along each line of sight. Recall that we are performing fits only on polarized data; including Stokes I data as well would require further model assumptions about how the polarized and unpolarized SEDs are related, which we leave to future work.

\begin{figure*}
\includegraphics[width=1.8\columnwidth]{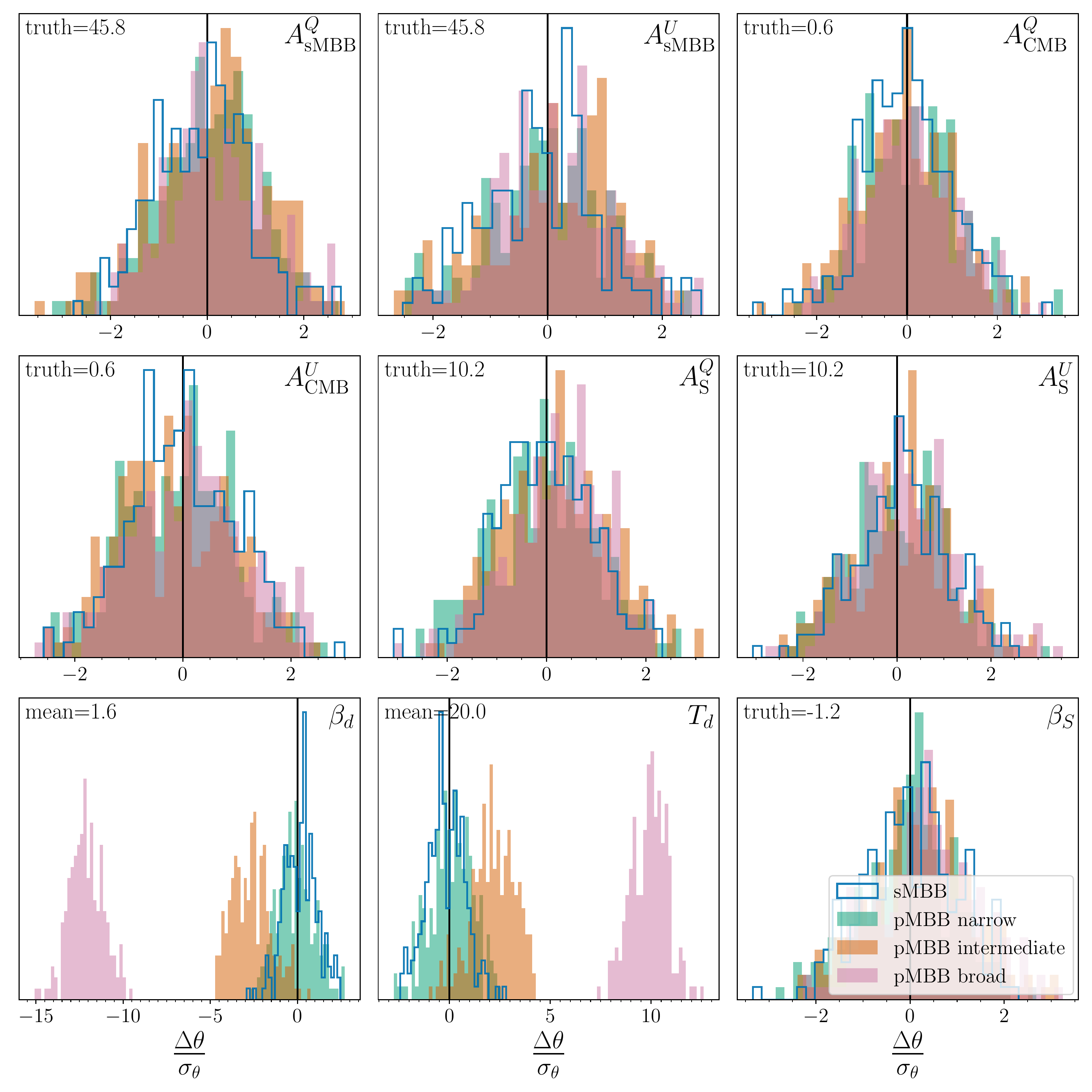}
\caption{The error-normalized bias values (see Eq.~\ref{eq:errnormbias}) of sMBB model parameters over N=200 noise realizations, for the four pMBB dust models considered. The input value (or input mean, for pMBB) is shown as the `truth' value in the upper left of each panel. Vertical black lines denote unbiased measurements.}
\label{fig:pMBB_biases}
\end{figure*}

\begin{figure*}
\centering
\includegraphics[width=1.78\columnwidth]{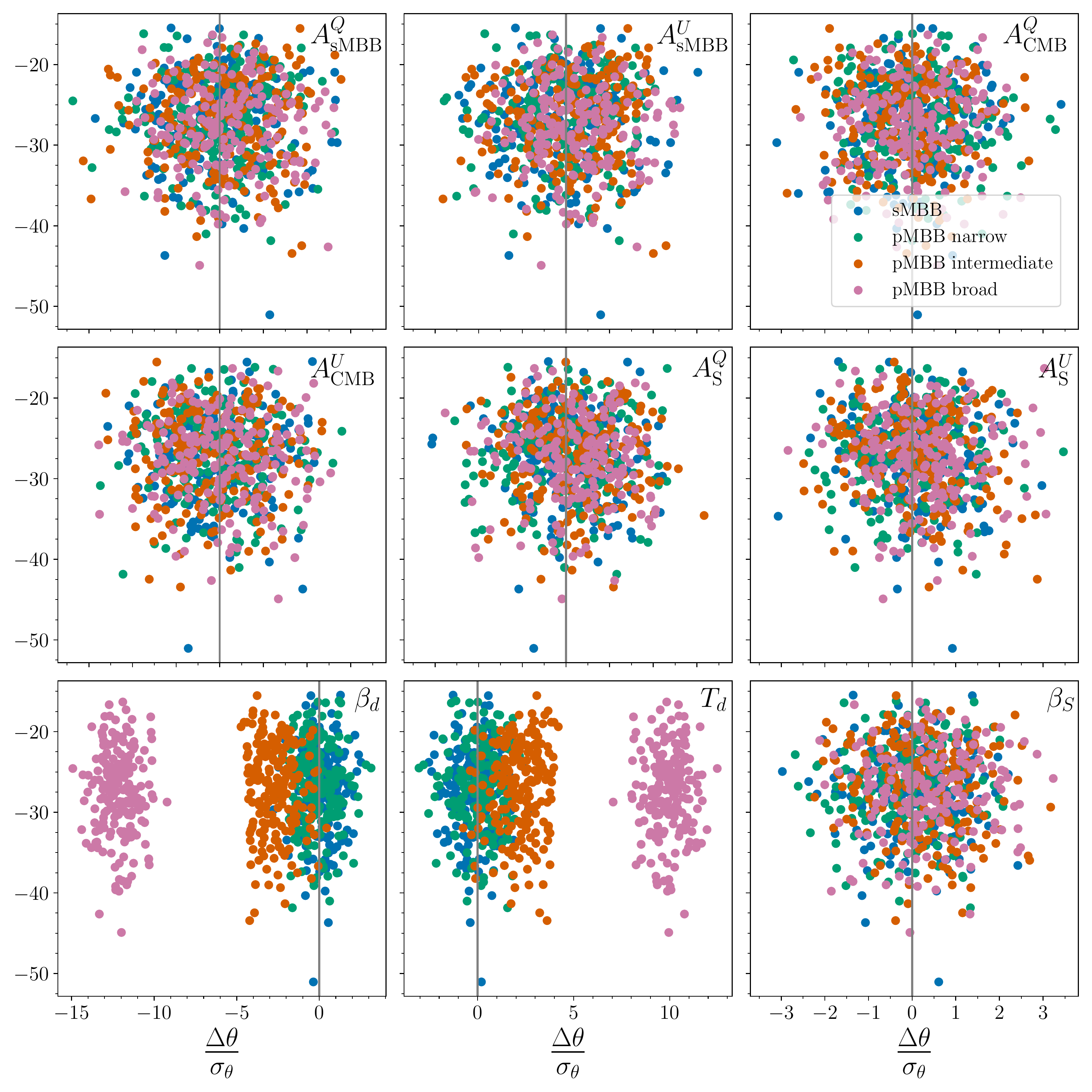}
\caption{Error-normalized bias values of the sMBB model parameters (x-axis) corresponding to the median value of the log posterior (y-axis), for the four dust models considered. For each input model, the values and spread of the log posterior median values are roughly the same, implying that the sMBB model produces equally good fits to the data in each case.}
\label{fig:pMBB_logp}
\end{figure*}

We considered four cases for the underlying (input) dust model. The first is simply an sMBB model, which we used as a check on our fitting procedure.  We then selected three options for the pMBB model, centered on the same mean value, but with varying widths for the $\beta_d$ and $T_d$ distributions: narrow, intermediate, and broad. The values of the widths of the distributions were summarized in Table~\ref{table:distributions}.

Fig.~\ref{fig:pMBB_biases} shows the error-normalized bias for the recovered best-fit parameters (including sMBB, CMB, and synchrotron parameters) in each of the four cases, for fits to each of 200 noise realizations per model. It can be seen that fitting an sMBB model to data that truly contains an sMBB model recovers unbiased estimates of all of the parameters on average, as expected. The same is largely true for the narrow pMBB model too, with no significant biases seen for any of the parameters.

The picture starts to change for the intermediate pMBB model however; while most parameters remain unbiased on average (including the CMB polarization amplitudes), the recovered dust temperature and spectral index parameters are significantly biased (by approximately $3\sigma$ and $2\sigma$ respectively). This is a sign that that the sMBB parameter fits are no longer representative of the properties of the underlying dust cloud population. The bias is even more pronounced for the broad pMBB case, where on average a $10\sigma$ discrepancy from the mean dust temperature is observed, and a $12\sigma$ discrepancy from the mean dust spectral index. Notably, $T_d$ and $\beta_d$ are inversely biased, which may be an artefact of the well-known degeneracy between these two parameters \citep{2009ApJ...696..676S}.

Next, we consider whether these biases would be detectable, in the sense that fitting the wrong model to the data could produce a noticeably poor goodness of fit. Fig.~\ref{fig:pMBB_logp} plots the maximum log-posterior value (i.e., $\log p$ for the best-fit model) for each realization against the error-normalized bias. There are no significant differences in the goodness of fit between the different cases, despite the significant biases in the recovered parameters, suggesting that an analyst would be unable to distinguish between the four scenarios by fitting sMBB models to the data alone.

\begin{figure*}
\includegraphics[width=1.8\columnwidth]{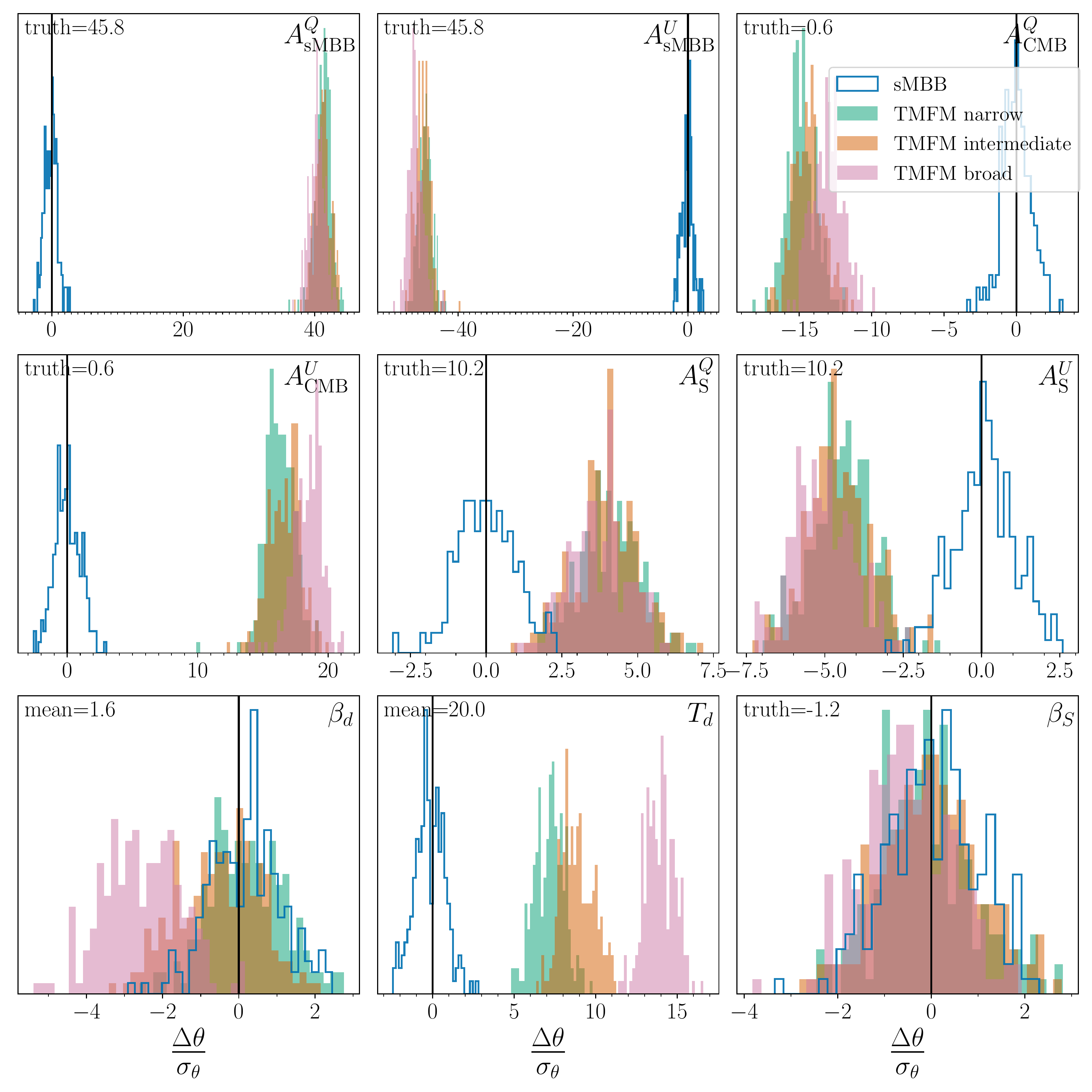}
\caption{Error-normalized bias values of sMBB model parameters over N=200 noise realizations, for the three widths of TMFM models considered.  Of particular note is the strong biasing of the CMB amplitudes in both polarization channels.}
\label{fig:TMFM_biases}
\end{figure*}

\subsection{Fitting the sMBB model to TMFM data}
\label{subsec:TMFM}

Next, we repeat our analysis but for the case of an sMBB model being fitted to data with an underlying TMFM dust model, i.e., now allowing significant polarization-dependent effects. Results for this model are shown as histograms in Fig.~\ref{fig:TMFM_biases} corresponding to three different values of the shape parameter, $\kappa$ (these three values again representing a narrow, intermediate, and broad case).
In sharp contrast to the results for the pMBB models, the best-fit CMB amplitudes are now significantly biased on average, and in fact only the dust amplitude and synchrotron spectral index do not exhibit strong biases. As expected, the biases become more pronounced as the size of $\kappa$ increases.  The size of the bias is also more strongly dependent on the specific noise realisation than for the pMBB models.  

\begin{figure*}
\includegraphics[width=1.95\columnwidth]{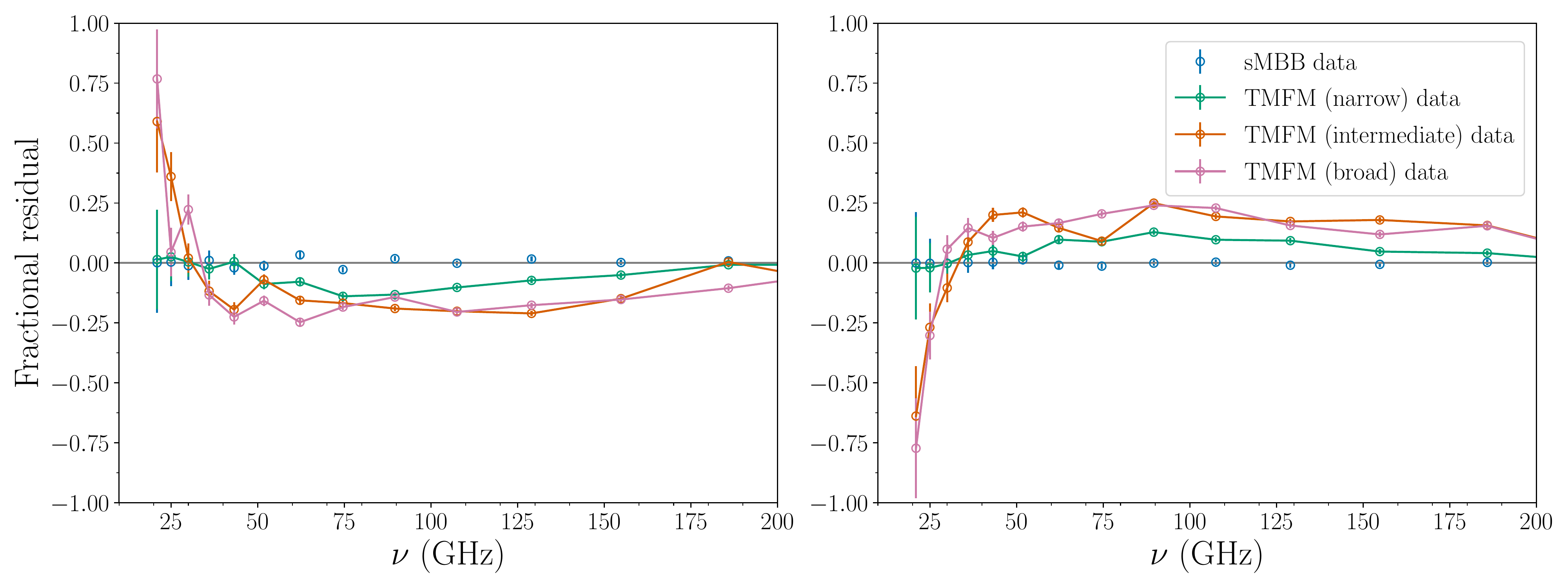}
\caption{Fractional residuals of the Q (left) and U (right) polarization channels, for the three different distribution widths for the TMFM model. Only the lower frequencies are shown to zoom in on the region of interest; a clear residual is seen at the lowest frequencies.}
\label{fig:residuals}
\end{figure*}

However, in this case the standard goodness of fit metric (i.e., maximum log-posterior values) is very low, indicating that the sMBB model is identified as giving a very poor fit to the data assuming PICO-like noise characteristics. Therefore these biases would be easily detectable. It is perhaps unsurprising that the sMBB model is not a good fit in this case, as it does not allow differences in SED between the different polarization channels, while the Stokes Q and U SEDs can be quite different for TMFM, as seen from the significant low-frequency residual shown in Fig.~\ref{fig:residuals}. 

\begin{figure*}
\centering
\includegraphics[width=1.82\columnwidth]{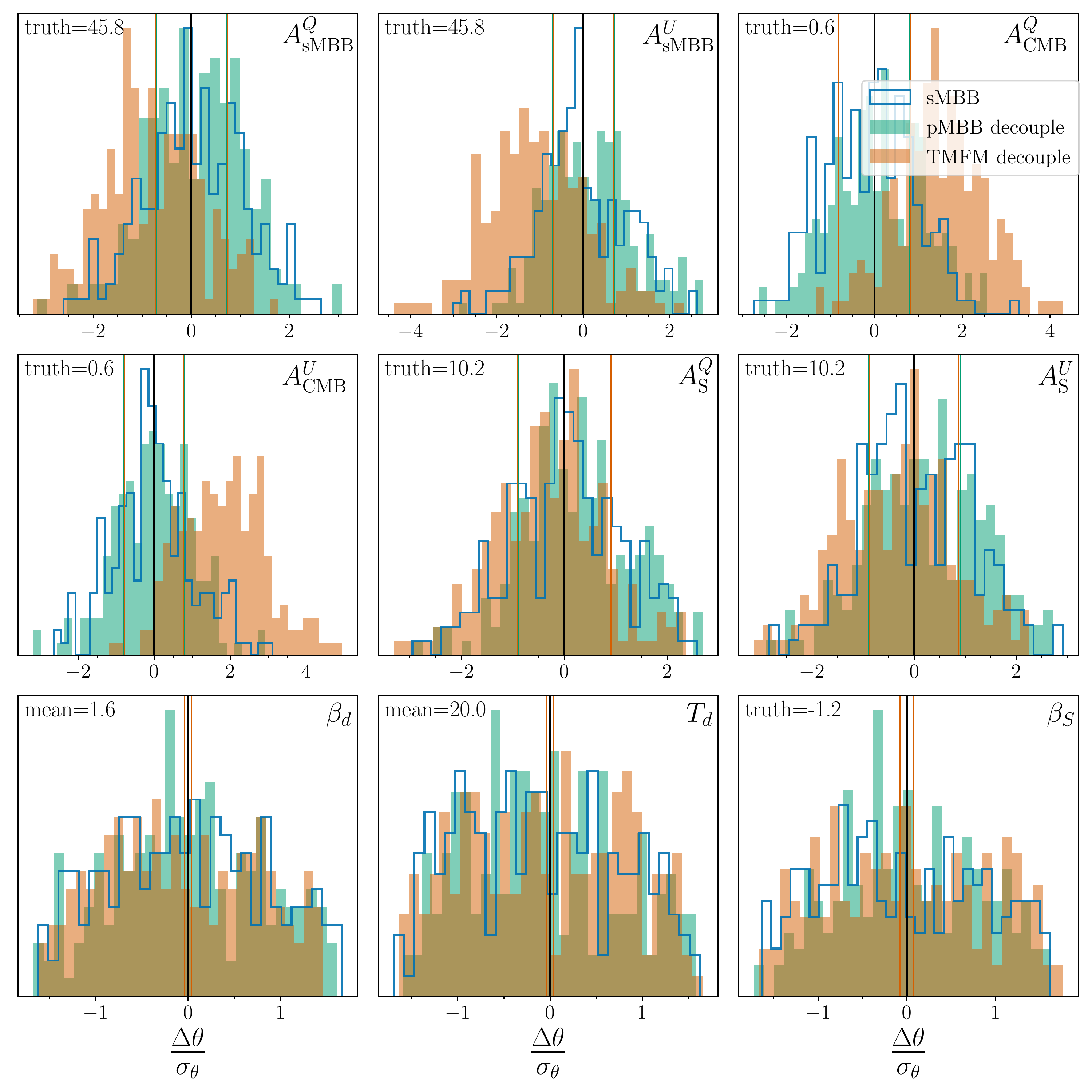}
\caption{Error-normalized bias values of sMBB model parameters over N=200 noise realizations, for one illustrative decoupled case for each model (sMBB, pMBB, and TMFM). The broadest distributions for both probabilistic models were used. Decoupling the two polarization channels greatly reduces the bias on the recovered parameters, at the expense of increasing the estimated uncertainties. For comparison, the $1 \sigma$ estimated errors from a non-decoupled MCMC run of each model are overlaid on each histogram (colored vertical lines in each panel). While each model's errors are overlaid, they are too close together to be easily distinguishable. The marginal distributions in the non-decoupled case are much narrower (the parameters are much better constrained).}
\label{fig:decouple_hists}
\end{figure*}

\begin{figure*}
\includegraphics[width=1.8\columnwidth]{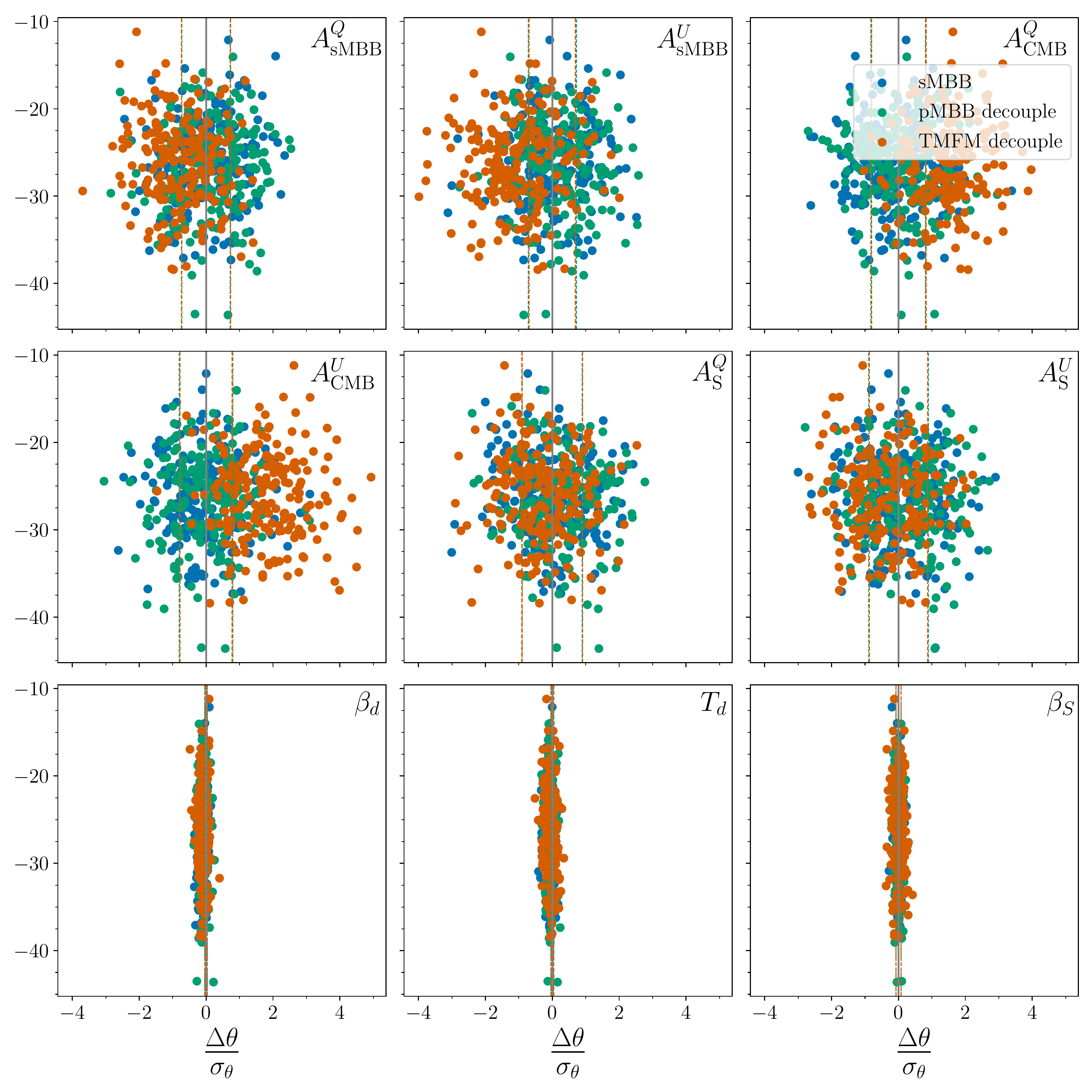}
\caption{Error-normalized bias values of sMBB model parameters (x-axis) versus the median value of the log posterior (y-axis), for the same illustrative decoupled cases for each model as shown in Fig.~\ref{fig:decouple_hists}.}
\label{fig:decouple_logp}
\end{figure*}

Because of the decorrelation between the Stokes Q and U channels in the TMFM model, we investigate whether it may be more practical to fit the two polarization channels independently. We performed the same analysis as for the above cases, but this time fitting the Q and U channels separately. The results are shown in Figs.~\ref{fig:decouple_hists} and \ref{fig:decouple_logp}. In this case, the results are not biased, however the resultant uncertainties are considerably larger. This is in line with what is expected from introducing a more flexible model---there is more freedom to fit the different SEDs in the Q and U channels, but more parameters to fit in total, which increases the uncertainty on all parameters across the board.

We have also checked the median log-posterior values for each model, in analogy to Fig.~\ref{fig:pMBB_logp}. We again found that the log-posterior values are consistent across all cases, with a reasonable goodness of fit for each of them; hence, studying the goodness of fit would not allow an analyst to distinguish between these scenarios.

These results, coupled with recent detections of frequency decorrelation related to the dust polarization angle in {\it Planck} data \citep{pelgrims2021evidence, Ritacco2022}, illustrate the importance of testing component separation methodologies with simulations of Galactic emission that include line of sight frequency decorrelation. Such models include the 3D ``layer'' model of \citet{MKD}, those based on the TIGRESS 3D magnetohydrodynamic simulations developed by \citet{Kim2019}, and the 3D filament model of \citet{HerviasCaimapo2022}.

\subsection{Inferring the hyperparameters of the dust parameter distributions}
\label{subsec:hyperparams}

We now turn to the question of whether the hyperparameters of the dust cloud parameter distributions themselves can be recovered from the data. This would be of significant interest in understanding the structure of the dusty ISM for example, as well as informing the construction of more realistic dust models for foreground removal.  To this end, we first attempted to perform the same MCMC analysis outlined above, but now fitting the hyperparameters of the pMBB model instead of the sMBB parameters.

\begin{figure*}
\includegraphics[width=2.15\columnwidth]{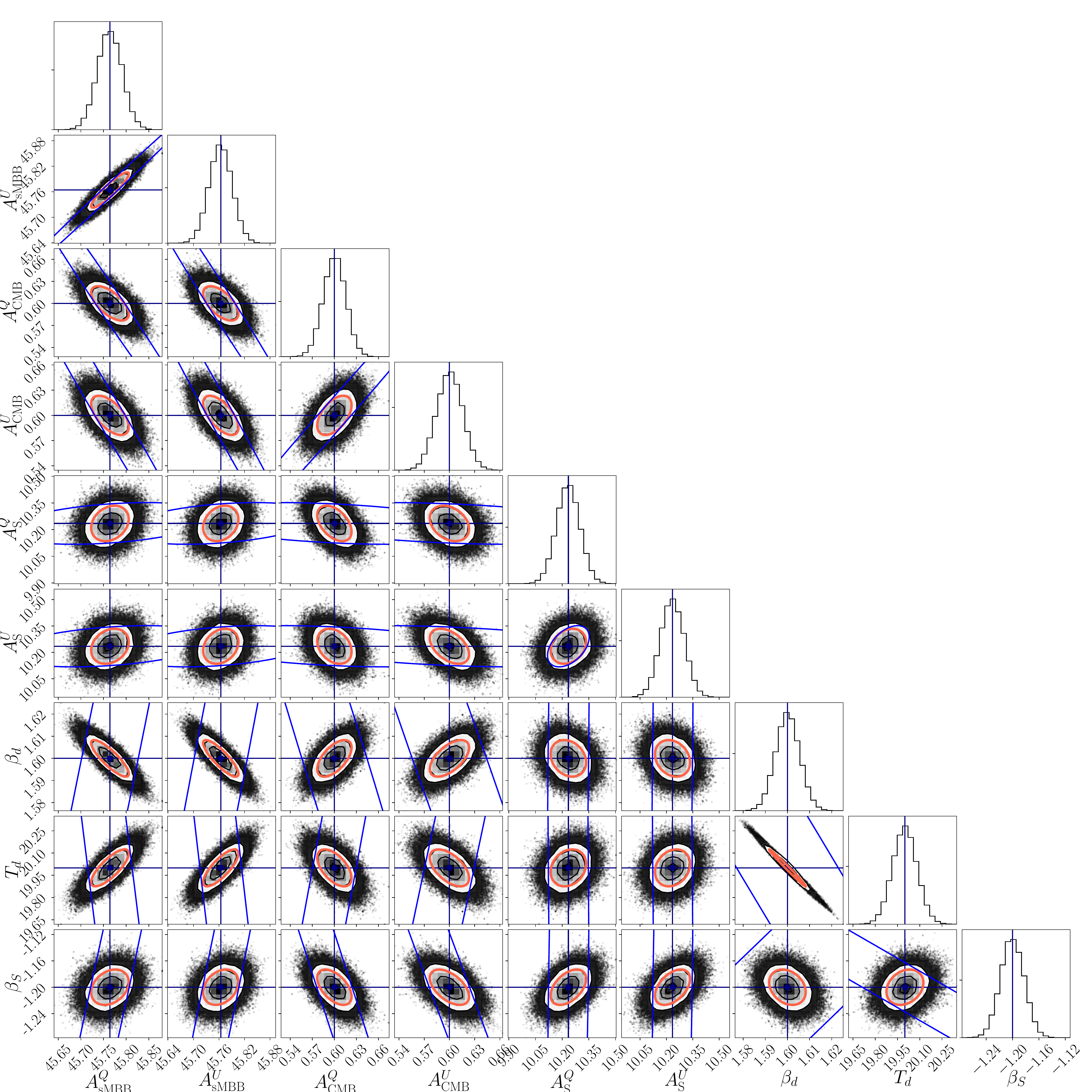}
\caption{A corner plot of one MCMC run, fitting the sMBB model to noiseless sMBB data (for noiseless data, we expect the best-fit model to be the true input model if the fitting procedure is working correctly).  Fisher confidence intervals for the sMBB model (red) and pMBB model (dark blue), using the same simulated sMBB model as the fiducial model of the data, are overlaid on the contours. Note the substantial degradation of all of the constraints once the means and widths of the dust parameter distributions are marginalized in the pMBB case. The vertical and horizontal lines (light blue) show the positions of the true input parameters (which are the same as the fiducial parameters for the Fisher analysis). All quantities have the same dimensions as in Table~\ref{table:priors}.}
\label{fig:corner_fisher}
\end{figure*}

We found that fitting the pMBB model suffered from convergence issues for data simulated from narrow pdf distributions (i.e., the sMBB, and pMBB narrow models).  The MCMC chains for the the wider distributions (the pMBB intermediate and broad models) were able to converge, and so we could study the recoverability of the hyperparameters in these cases. The confidence intervals derived from the MCMC samples were much smaller than those calculated from the Fisher matrix formalism however (see below), which implies that the MCMC is not fully exploring the relevant portions of the posterior distribution.\footnote{An exploration of ways to improve the convergence of the MCMC fitting to the pMBB model was conducted. Techniques that were tried included a suite a different initialisation positions for the walkers, running noiseless simulations, and fitting modified functions of the hyperparameters (e.g., as a proxy for alternative prior distributions), which appear to be the cause of the convergence issues.  While these various changes did improve convergence, they ultimately didn't prove successful enough to confidently determine the constraints on the hyperparameters.} A comparison between the MCMC and Fisher approaches is shown in Fig.~\ref{fig:corner_fisher}.

As an alternative to the MCMC-based method, we also performed Fisher matrix forecasts to understand how well the statistical dust model hyperparameters could be constrained by future CMB experiments. While approximate, Fisher forecasting has the advantage of being fast and comparatively simple, allowing forecasts for many different experimental configurations to be studied en masse. While the forecasts are by nature optimistic (they constitute a `best-case' estimate of uncertainty given a particular experimental configuration), and do not account for non-Gaussianities in posteriors and other such complications, they are a useful method of comparison.

For the single pixel SED-fitting problem, the Fisher matrix can be calculated as
\be
F_{ij} = \sum_\nu \left . \frac{1}{\sigma_\nu^2} \frac{\partial m(\vec{\theta})}{\partial \theta_i} \frac{\partial m(\vec{\theta})}{\partial \theta_j} \right |_{\vec{\theta}_0},
\ee
where $\sigma_\nu$ is the noise rms per frequency band (assumed uncorrelated), $m(\vec{\theta}) = \sum_k A_k f_k(\nu; \vec{\theta})$ is the signal model along the line of sight, summed over all components $k$, and $\vec{\theta}_0$ is a set of fiducial model parameters. As in our MCMC studies, we assume that all frequency bands have been smoothed to the same angular resolution, and that there is a $\delta$-function bandpass response. 
The predicted parameter covariance matrix can then be approximated by inverting the Fisher matrix, $C_{ij} \approx (F^{-1})_{ij}$, where the elements of $C_{ij}$ are the expected covariance of parameters $\theta_i$ and $\theta_j$ marginalized over all other parameters. The free parameters included in the Fisher matrix are the same as those included in the MCMC fits, i.e., we marginalize over the CMB, synchrotron, and dust amplitude parameters, as well as the synchrotron spectral index.

The Fisher formalism was first applied to the case of fitting an sMBB model to data generated from an underlying sMBB model. All derivatives for both this case and the pMBB model were calculated numerically using a simple finite difference method. The estimated 68\% CL result is overlaid onto an illustrative MCMC run in Fig.~\ref{fig:corner_fisher}, where the input data vector was a noiseless sMBB SED for the same set of fiducial model parameters. The two different methodologies are in good agreement with one another, which provides a check on our Fisher matrix machinery.

\begin{table}
\centering 
\def\arraystretch{1.15} 
\begin{tabular}{ l | S[table-format=3.2] S[table-format=3.3] | S[table-format=3.2] S[table-format=3.3] }
 \hline
{\bf Dust model} & {SNR($\bar{\beta}$)} & {SNR($\sigma_{\beta}$)} & {SNR($\overline{T}_d$)} & {SNR($\sigma_{T}$)} \\
 \hline
 pMBB narrow        & 5.68   & 0.002   & 3.08 & 0.004 \\
 pMBB inter.  & 4.91   & 0.045   & 2.53 & 0.101   \\
 pMBB broad         & 13.4   & 0.414   & 5.38 & 3.61    \\
 \hline
\end{tabular}
\caption{Forecast signal-to-noise ratios for the pMBB hyperparameters, derived from the Fisher matrix analysis.} 
\label{table:forecasts} 
\end{table}

Next, the Fisher matrix was calculated for the scenarios of fitting the pMBB model hyperparameters to the narrow, intermediate, and broad pMBB fiducial models in a single pixel. The results are shown in Table~\ref{table:forecasts} as signal to noise ratios, i.e., the ratio of the fiducial value to the forecast $1\sigma$ error for each parameter. In general, we found that the means of the $\beta$ and $T_d$ distributions are measurable, but the widths (standard deviations) are not, at least for a single pixel. The broad distribution is the easiest to characterise, with our forecasts predicting good measurements of $\bar{\beta}$ and $\overline{T}_d$ (SNRs of 13 and 5 respectively), and a $3.6\sigma$ measurement of $\sigma_T$. The width of the spectral index distribution, $\sigma_\beta$, is not measurable however, with a predicted SNR of only $0.4$. The picture for the intermediate and narrow distributions is less encouraging, with very poor constraints on both width parameters, but reasonable (${\rm SNR \sim 3-6}$) measurements of the means of the distributions. Note that the means can actually be measured slightly better in the narrow case than in the intermediate case; this appears to be due to correlations between parameters changing as the constraints on the distribution width parameters improve significantly (most other parameter constraints, e.g., for the synchrotron and CMB parameters, are quite similar between the narrow and intermediate cases). 

Some additional context for these results is provided by Fig.~\ref{fig:frac_diff}, which shows the fractional difference of the pMBB SEDs with the reference sMBB model. The broad distribution is the only one that deviates substantially from sMBB at both high and low frequencies, which goes some way to explaining its better forecast constraints. It also has a larger deviation from sMBB at low frequencies, where the PICO mission has more bands, but the dust SED itself is lower in intensity \citep[not shown in the figure; the thermal dust polarized intensity becomes sub-dominant to other foregrounds at around 60~GHz;][]{Planck_foregrounds}. The narrow case is essentially degenerate with sMBB, so the distribution mean parameters $\bar{\beta}$ and $\overline{T}_d$ can be identified with the $\beta$ and $T_d$ parameters of sMBB. The intermediate case only deviates from sMBB significantly at low frequencies, with a deviation that is approximately five times smaller than the one for the broad distribution.

In terms of prospects for measuring the dust parameter distributions with future microwave experiments, the main implication of our Fisher analysis is that the distributions along individual lines of sight will probably only be characterisable if they are quite broad, i.e., if there is substantial variation in physical properties from dust cloud to dust cloud. This is because narrower parameter distributions result in SEDs that deviate comparatively little from simple single MBB models. While we have only looked at a few particular cases, and have not explored distributions with correlations between $\beta$ and $T_d$ for example, we did consider different shapes of distribution (see Sect.~\ref{subsec:stat_models}), finding little difference in the resulting SEDs. We therefore expect this conclusion to be reasonably robust. Indeed, as we showed in the previous section, the polarisation properties of the clouds seem to have a much stronger effect on the SEDs.

\section{Conclusions} \label{sec:conclusions}

In this paper, we studied a probabilistic thermal dust model based on the modified blackbody (MBB) model commonly used for component separation in CMB experiments. This `probabilistic MBB' (pMBB) model was intended to account for the variation of physical properties that occurs between the many individual dust clouds within the angular beam of a CMB experiment along each line of sight. We investigated the implications of fitting simulated data generated with this probabilistic model to both the simplistic single MBB (sMBB) model, and the pMBB model itself.

In Section \ref{sec:models} we described the single-population MBB dust model, in addition to introducing our statistical models, both the probabilistic MBB models, and a Turbulent Magnetic Field model (TMFM). The latter model incorporates line of sight frequency decorrelation effects that lead to different frequency dependence in Stokes Q versus U \citep{tassis2015searching}.

We found that even in the most extreme pMBB case considered (that of a broad distribution in the physical dust parameters), the CMB signal remains unbiased when fitting a simple sMBB model to the data. The fitted dust parameters are biased from the mean values of the pMBB distribution by several $\sigma$ however, which suggests that any physical conclusions drawn about the dust cloud population from the sMBB parameters could be misleading. The biases of these values appear to be inversely related, in that a lower fitted temperature $T_d$ is offset by a higher value of $\beta_d$. This reproduces the well-known dust temperature-spectral index degeneracy \citep[e.g.][]{2009ApJ...696..676S}.

When polarization effects exist, as illustrated by the TMFM model, we find that fitting an sMBB model does tend to lead to significant biases in most parameters, including the polarized CMB amplitudes. Essentially, the sMBB model is unable to absorb differences in the shapes of the SEDs between polarization channels that arise, particularly at lower frequencies. However, this also gives rise to poor goodness of fit statistics, which would make it relatively clear to an analyst that an sMBB model fit is inappropriate in this scenario. This is in contrast to the pMBB models, for which the goodness of fit statistics are equally good for all three pMBB data sets, and much the same as for the sMBB data. In other words, the pMBB data without polarization effects gives rise to biases in the dust model parameters that are hard to detect, whereas the TMFM data give rise to biases that are easy to detect.

As a further check, we also tried fitting `decoupled' sMBB models to the pMBB and TMFM scenarios, in which a different sMBB model was fitted per polarization. This largely mitigated the biases observed for both models, at the expense of broadening the marginal distributions of the recovered parameters as more parameters were being fitted in total. We note that the CMB amplitudes could remain biased by around $1\sigma$ on average for the TMFM model with the broadest distribution however (see Fig.~\ref{fig:decouple_hists}), at least for the PICO-like experimental setup we considered. 

Finally, we also considered the possibility of trying to recover the (hyper)parameters of the pMBB model from the data. Fitting the pMBB model directly using an MCMC approach proved to be difficult from a numerical standpoint, and we have deferred a solution to this issue to later work. Instead, we performed Fisher forecasts to understand how well the pMBB hyperparameters could in principle be constrained in each $1\,{\rm deg}^2$ pixel of an experiment like PICO, with frequency coverage over a wide range of the microwave spectrum. We found that all hyperparameters except the width of the spectral index distribution, $\sigma_\beta$, could be recovered with a reasonable signal-to-noise ratio ($\gtrsim 4$) in the case of a fiducial model with the broadest distribution. Only the means of the dust temperature and spectral index distributions were recoverable for models with narrow- and intermediate-width distributions however, with little prospect of measuring the widths of those distributions according to our forecasts. This places an interesting limitation on how well we may be able to infer the properties of the dust cloud distribution in the Milky Way from future CMB data alone; only if there is substantial variation from dust cloud to dust cloud along each line of sight will we be able to actually measure the properties of the distributions in a single sky pixel. Moment-based methods applied over large sky areas offer a promising way to go beyond this limitation and constrain the true distributions of physical parameters using upcoming polarization data \citep[e.g.,][]{chluba2017rethinking, Remazeilles_2021, Vacher_2022}.

In conclusion, we find that the single MBB model can reasonably be used for per-pixel foreground cleaning in CMB polarization experiments in the absence of depolarization effects, but that physical conclusions about the dust properties obtained from such fits should be treated with caution. Conversely, single MBB models should not be used in the presence of depolarization effects, as strong biases in both the recovered CMB and dust parameters can be obtained. This is partly, but not fully, ameliorated if separate sMBB models are fit to each of Stokes Q and U, at the cost of increased uncertainties on the recovered parameters.

As a final note, we point out that we have only considered relatively ad hoc models and distributions in this paper. Physical parametric models that more accurately describe the actual composition of the dust, and thus the associated statistical parameters, if they could be recovered, could yield information about the nature of dust in the ISM itself. The construction of sufficiently realistic physical models is an on-going process, and one that requires observations of dust not just at microwave frequencies, but across the electromagnetic spectrum.

{\bf Note added ---} Immediately prior to submission, a paper by \citet{2022arXiv220713109S} appeared that also studies the impact of line-of-sight variation in dust properties on recovery of the polarized CMB temperature fluctuations. Their study also uses PICO as the target instrument and considers a range of probability distributions for the dust temperature and spectral index. A key methodological difference is that we fit the sMBB and pMBB models directly, while they construct a moment expansion of the kind introduced by \citet{chluba2017rethinking}.

They find that models with $\sigma_T \gtrsim 1.6$~K and/or $\sigma_\beta \gtrsim 0.045$ can bias the scalar-tensor ratio $r$ measured from the polarized CMB if sMBB (single MBB) models are used to fit the dust foreground in each pixel. This is closest to our `pMBB intermediate' case ($\sigma_T = 2.0$~K and $\sigma_\beta = 0.10$), for which we found a small mean bias on the recovered polarized CMB amplitudes for a PICO-like experiment (1~deg$^2$ pixel size) of $\approx 0.12 \sigma$; see Table~\ref{table:biases}. For a typical value of the error on the measured CMB Q/U amplitude of $\approx 0.1~\mu{\rm K}$ (see Fig.~\ref{fig:corner_fisher}), this corresponds to a bias of $\Delta Q \approx \Delta U \approx 10$~nK, which is roughly a factor of $2$ larger than the corresponding results shown in Figs.~3 and 4 of \citet{2022arXiv220713109S}. This is a relatively small discrepancy, and may be caused by the fact that we have chosen $\sigma_T$ and $\sigma_\beta$ to be non-zero simultaneously; so, we believe this result to be approximately consistent with ours.

\citet{2022arXiv220713109S} also claim that the line-of-sight dust parameter distributions can be recovered successfully using the moment expansion, particularly as larger numbers of moments are included. Conversely, we found that the parameters were difficult to recover in our pMBB Fisher forecasts, except for in the pMBB broad case (see Table~\ref{table:forecasts}). Part of the reason for this discrepancy may be that \citet{2022arXiv220713109S} use the continuous, noise-free SED for their moment fitting, whereas we use PICO-like frequency bands with PICO-like noise. An exception is for the `transient heating model' (their Fig.~10), which does use PICO-like instrumental properties. This model's (relatively broad) distribution in dust temperature is recovered to within roughly a factor of two, but with a somewhat poor fit to the shape of the distribution, even with six moments fitted. While we are unable to make a direct comparison of these results, both papers are consistent in finding that broad distributions can be characterized observationally with a PICO-like experiment.

\section*{Acknowledgements}

We are grateful to A.~Liu for useful discussions. This result is part of a project that has received funding from the European Research Council (ERC) under the European Union's Horizon 2020 research and innovation programme (Grant agreement No. 948764; PB). PB acknowledges support from STFC Grant ST/T000341/1. BSH acknowledges support from the NASA TCAN grant No.~NNH17ZDA001N-TCAN. 
This research was enabled in part by support provided by Calcul Qu\'{e}bec (\url{calculquebec.ca}) and the Digital Research Alliance of Canada (\url{alliancecan.ca}).

\section*{Data availability}

Code and associated data files used in this analysis are available from \url{http://philbull.com/singlepixel/}. Monte Carlo outputs can be regenerated by using these scripts/data.

\balance



\bibliographystyle{mnras}
\bibliography{article_bib} 

\bsp	
\label{lastpage}
\end{document}